# Pop III *i*-process Nucleosynthesis and the Elemental Abundances of SMSS J0313-6708 the Most Iron-Poor Stars

O. Clarkson,[1,3,†]⋆ F. Herwig,[1,3,†] M. Pignatari,[2,3,†]

[1]*Department of Physics & Astronomy, University of Victoria, P.O. Box 3055 Victoria, B.C., V8W 3P6, Canada*
[2]*E.A. Milne Centre for Astrophysics, University of Hull, HU6 7RX, United Kingdom*
[3]*Joint Institute for Nuclear Astrophysics, Center for the Evolution of the Elements, Michigan State University,*
*640 South Shaw Lane, East Lansing, MI 48824, USA*
[†] *NuGrid Collaboration*

## ABSTRACT

We have investigated a highly energetic H-ingestion event during shell He burning leading to H-burning luminosities of $\log(L_H/L_\odot) \sim 13$ in a $45 M_\odot$ Pop III massive stellar model. In order to track the nucleosynthesis which may occur in such an event, we run a series of single-zone nucleosynthesis models for typical conditions found in the stellar evolution model. Such nucleosynthesis conditions may lead to *i*-process neutron densities of up to $\sim 10^{13} \mathrm{cm}^{-3}$. The resulting simulation abundance pattern, where Mg comes from He burning and Ca from the *i* process, agrees with the general observed pattern of the most iron-poor star currently known, SMSS J031300.36-670839.3. However, Na is also efficiently produced in these *i* process conditions, and the prediction exceeds observations by $\sim 2.5$dex. While this probably rules out this model for SMSS J031300.36-670839.3, the typical *i*-process signature of combined He burning and *i* process of higher than solar [Na/Mg], [Mg/Al] and low [Ca/Mg] is reproducing abundance features of the two next most iron-poor stars HE 1017-5240 and HE 1327-2326 very well. The *i* process does not reach Fe which would have to come from a low level of additional enrichment. *i* process in hyper-metal poor or Pop III massive stars may be able to explain certain abundance patterns observed in some of the most-metal poor CEMP-no stars.

## 1 INTRODUCTION

Pop III stars produced the first elements heavier than those created in the Big Bang and polluted the surrounding pristine gas (Nomoto et al. 2013). The most metal-poor stars we observe today may be the most direct descendants, or at least carry the most distinct signatures, of Pop III stars and therefore become a powerful diagnostic in our study of early cosmic chemical evolution (Frebel & Norris 2015) .

Of the most iron-poor stars, the majority are classified as carbon enhanced metal poor -no (CEMP-no) (Beers & Christlieb 2005). SMSS J031300.36-670839.3 (hereafter SMSS J0313-6708, Keller et al. 2014), is the most iron-poor star identified at present, with [Fe/H] $\le -6.53$, (Nordlander et al. 2017). Li, C, Mg and Ca have been measured and there are upper limits on several other elements. HE 1327-2326 (Frebel et al. 2006, 2008) and HE 1017-5240 (Christlieb et al. 2004) are the next two most iron poor stars known with [Fe/H] -5.96 and -5.3, respectively.

Low-energy or faint supernovae with strong fallback would have either very little or no nucleosynthetic contribution from the supernova explosion (Keller et al. 2014; Takahashi et al. 2014; Marassi et al. 2014). However, many the best fit models proposed for CEMP-no stars are within the mass range where Pop III and the lowest-metallicity stars are expected to collapse directly into black holes with no supernova explosion (Heger et al. 2003). Choplin et al. (2016) proposed that progenitors of such stars may have been massive, rapidly rotating, Pop III stars. Takahashi et al. (2014) found rotating Pop III models less favourable in reproducing the abundances of SMSS J0313-6708 than non-rotating models, but preferable for HE 1017-5240 and HE 1327-2326.

Overall, there is currently no clear consensus on either the production site or mechanism which would explain the observed abundances of SMSS J0313-6708, apart from the zero-metallicity nature of the progenitor. Here we are proposing a new nucleosynthesis mechanism that can operate in Pop III as well as hyper metal-poor massive stars.

⋆ E-mail: oclark01@uvic.ca

H-ingestion events into the He-burning core or shell in Pop III and low-metallicity massive stars have been reported based on models with different stellar evolution codes and physics assumptions (Marigo et al. 2001; Heger & Woosley 2010; Limongi & Chieffi 2012; Takahashi et al. 2014; Ritter et al. 2017). The events affect the structure, evolution and nucleosynthetic yields of Pop III stellar models, but fundamental questions concerning the occurrence conditions and properties remain unanswered.

We investigate the possibility that nucleosynthesis patterns of CEMP-no stars SMSS J0313-6708, HE 1017-5240, and HE 1327-2326 contain the nucleosynthesis signatures of convective-reactive H-ingestion events. Such events would amount to a light-element version of the $i$ process, a neutron capture process with neutron densities in the range $10^{13} - 10^{15}\,\mathrm{cm}^{-3}$, that is activated in convective-reactive, combined H and He-burning events (Cowan & Rose 1977; Dardelet et al. 2014; Herwig et al. 2011; Hampel et al. 2016). We propose that this event may produce sufficient energy to expel a portion of the H/He convective-reactive layer of the star as discussed by Jones et al. (2016).

Section 2 describes the stellar evolution models, Section 3 the nucleosynthesis simulations and comparison with observations, and in Section 4 we conclude.

## 2   1D STELLAR EVOLUTION MODEL

We use the MESA stellar evolution code (Paxton et al. 2015, rev. 8118). Assumptions include the Ledoux criterion and semiconvection (Langer et al. 1985) with efficiency parameter $\alpha = 0.5$. The custom nuclear network includes 82 species with $A = 1 - 58$. We neglect stellar mass loss because Pop III stars likely have inefficient line-driven winds (Krtička & Kubát 2006). We ignore the effects of rotation.

The abundances or upper limits for Fe in the most iron-poor stars investigated here require the assumption of a low-energy supernova with strong fallback and little mixing. These stars are inconsistent with nuclear production of pair-instability supernovae making masses $\sim 140 - 260\,\mathrm{M}_\odot$ improbable (Keller et al. 2014). We have chosen an initial mass of $45\mathrm{M}_\odot$ which is expected to collapse into a black hole without SN explosion (Heger et al. 2003). We have explored other masses and they harbour similar thermodynamic conditions (Section 3). At this point we consider the stellar evolution simulations as guide for our nucleosynthesis calculations rather than a definitive solution.

We initialize with Big Bang abundances of Cyburt et al. (2016). The main-sequence and core-He burning phases follow previous descriptions closely (e.g. Marigo et al. 2001; Limongi & Chieffi 2012). The time evolution of this model is shown in Fig. 1. Soon after the exhaustion of core He, a convective He-burning shell develops. After $\approx 2.5 \times 10^3$ yr the He and H burning layer begin to interact and exchange material. H entering the He convection zone leads to energy generation at the interface of the layers. The entropy difference between the two layers before they come into contact is $\Delta S/N_\mathrm{A} k_\mathrm{B} \approx 7.5$, a factor of about 7 − 8 less than in corresponding models of solar metallicity. In the model, mixing occurs intermittently between the H shell and the He shell below, separated by a radiative layer with a radial extent of $2.7 \lambda_P$ from the base of the He shell. Just prior to the ingestion event the entropy difference has been reduced to $\Delta S/N_\mathrm{A} k_\mathrm{B} \approx 5$. From here H and a small amount of its associated burning products are mixed downward into the partially-burned He layer below. Nuclear energy production increases within minutes (Fig. 1), and the burning of H creates a split in the He-shell, similar to Herwig et al. (2011). 3D simulations are only starting to investigate this process with the necessary numerical effort (Herwig et al. 2014), but already show that violent, global instabilities are possible. The 3D behaviour is expected to be fundamentally different compared to what is seen in 1D stellar evolution models.

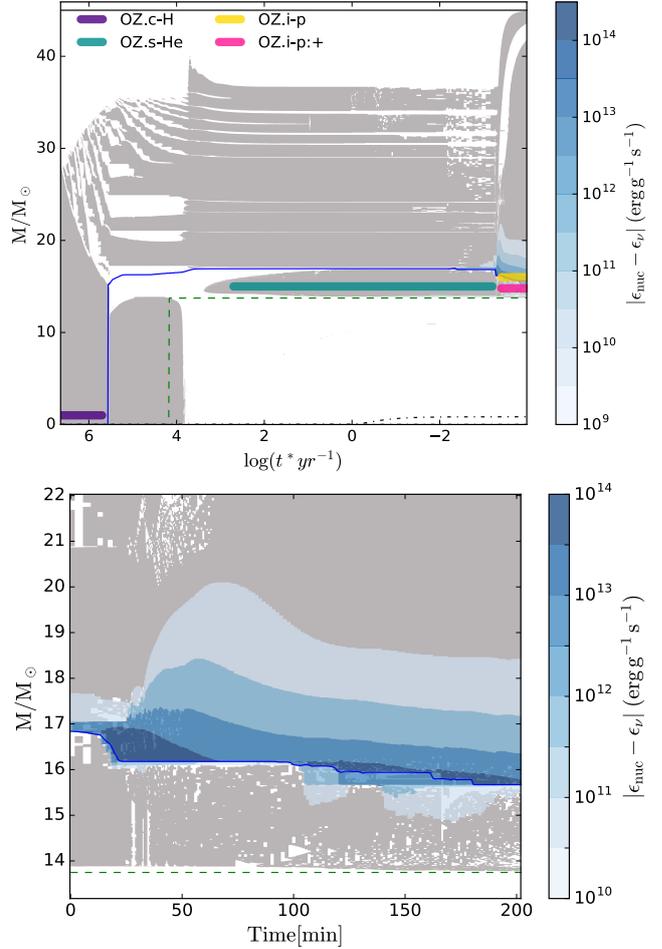

**Figure 1.** **Upper panel:** Evolution of convection zones (grey), nuclear energy generation (blue contours) and the H- (solid blue) and He- (green dashed) and C-free (black dashed) cores for the $45\mathrm{M}_\odot$ stellar evolution model. Purple, teal, yellow and pink lines schematically illustrate the regimes where single-zone calculations are preformed (Section 3). **Lower panel:** Zoom-in of H-ingestion event shown in linear time with $t = 0$ the beginning of the event.

During this event, energy generation is dominated by $^{13}\mathrm{C}(\alpha, n)^{16}\mathrm{O}$. The luminosity in this region reaches $\log(L_\mathrm{H}/\mathrm{L}_\odot) \sim 13$. Following the approach of Jones et al. (2016) we calculate the maximum value $H = \epsilon_\mathrm{nuc} \tau_\mathrm{conv}/E_\mathrm{int} \approx 0.26$, where $\epsilon_\mathrm{nuc}$ is the specific energy generation rate of nuclear reactions, $E_\mathrm{int}$ is the specific internal energy, both measured within the upper portion of the split convection zone $(\sim 15.5 - 17.0 \mathrm{M}_\odot$, Fig. 1). $\tau_{\mathrm{conv}}$ is the a convective timescale. Thus, a significant fraction of the binding energy of the layer is being deposited into this region of the star on a single mixing timescale, and, following the arguments of Jones et al. (2016), suggests a dynamic response that violate the assumptions of mixing length theory (MLT). MLT approximates convection through spatial and time averages over many convective turnover time scales and is applicable in non-dynamic, quiescent burning regimes. In the simulation presented here, it is expected that the large amount of energy generated from nuclear reactions will feedback into the flow in such a way that the MLT assumptions break down.

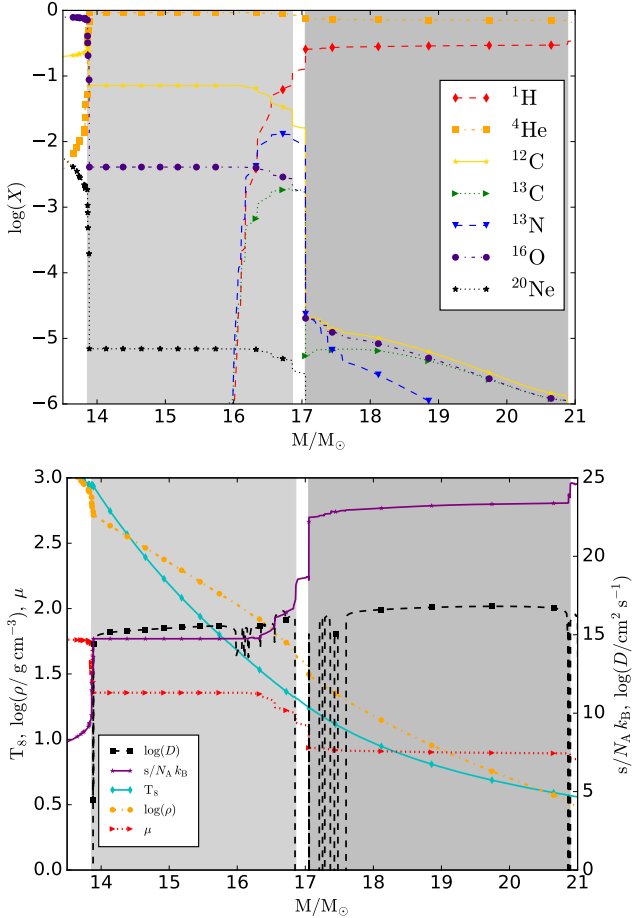

**Figure 2.** Grey areas show convective H and He-burning regions just before H begins mixing into to He shell. Temperature, density, mean molecular weight, entropy and diffusion coefficient for convective mixing shown 9 min later.

Towards the end of the lives of some massive stars, nonterminal, discrete mass loss events are detected as supernova type IIn or supernova imposters (see Section 4 of Smith et al. (2014). Arnett et al. (2014) suggest that these types of mass ejection events require 3D modelling, as MLT assumes a steady state whereas the late stages of massive stellar evolution are likely highly dynamic. 3D calculations with full $4\pi$ geometry performed by Herwig et al. (2014) demonstrate that under convective-reactive conditions, severe departures from spherical symmetry can occur. We hypothesise that something akin to a GOSH, or Global Oscillation of Shell H-ingestion, (Herwig et al. (2014)) may occur in the model presented here. This must be verified by 3D hydrodynamic simulations. 1D calculations of similar H-ingestion events into the He burning shell in Super-AGB stars have been presented in Jones et al. (2016) where it is argued that H-ingestion events with similar H numbers could launch such outbursts. If so, even a relatively small amount of *i*-process enriched material could be ejected and enrich the surrounding ISM where then a second generation star forms, possibly with distinct abundance signatures. For now, the precise details of such a mechanism are beyond the scope of this letter.

## 3 NUCLEOSYNTHESIS CALCULATIONS

To study the nucleosynthesis in the H/He convective-reactive environment we employ separate nucleosynthesis calculations using the NuGrid single-zone PPN code (Pignatari et al. 2016). The single-zone method was chosen over multi-zone simulations, because the large H number of the convective-reactive event suggests that the 1D modelling assumptions of convection break down. Instead, one-zone simulations—although constituting a further simplification, allow studying the nucleosynthesis that may be possible in this event in isolation. The dynamic NuGrid network includes as required up to 5234 isotopes with associated rates from JINA Reaclib V1.1 (Cyburt et al. 2010) and other additional sources (see Pignatari et al. 2016). The general strategy is to approximate the nucleosynthesis through a series of three one-zone calculations which start with H burning followed by He burning and finally, we add in the last step a small amount of H to the partially completed He-burning nucleosynthesis calculation to estimate the nucleosynthesis due to H ingestion. The thermodynamic parameters for each of these three steps (Table 1) are taken to represent the conditions found in the stellar evolution simulation, as shown for the onset of the H-ingestion phase in Fig. 2.

Each of the one-zone simulation steps contributes to the final abundance distribution (Fig. 4). The H-burning simulation (OZ.c-H) starts with the same Big Bang abundances as the stellar evolution model. The OZ.c-H is evolved until it reaches the same CNO abundances as the stellar evolution model does at the end of H-core burning, which requires less time in the one-zone simulations because it does not include convective mixing. The output from this burning stage is used to initialize the He-burning one-zone run. Two separate cases are considered (Table 1). OZ.s-He very closely follows the relatively small amount of He shell burning found in our stellar evolution model up to the point when the H and He shells start to interact.

**Table 1.** Parameters for single-zone PPN calculations. Row 4 contains a single run with three output times.

| Run ID | Burning phase | T ($10^8$K) | $\rho$ (g cm$^{-3}$) | $\Delta t$ (yrs) |
|---|---|---|---|---|
| OZ.c-H | Core H | 1.25 | 93.33 | $2.21 \times 10^4$ |
| OZ.s-He | Shell He | 2.6 | 330 | $1.28 \times 10^2$ |
| OZ.s-He:+[†] | Shell He | 2.95 | 487.1 | $4.45 \times 10^2$ |
| OZ.i-p:t1,2,3 | H-ingest. | 2.0 | 191 | $1,2,5 \times 10^{-2}$ |
| OZ.i-p:+ | H-ingest. | 2.41 | 315.4 | $3.44 \times 10^{-2}$ |

[†]Single zone run representing more efficient and complete He burning. For details see section 3.

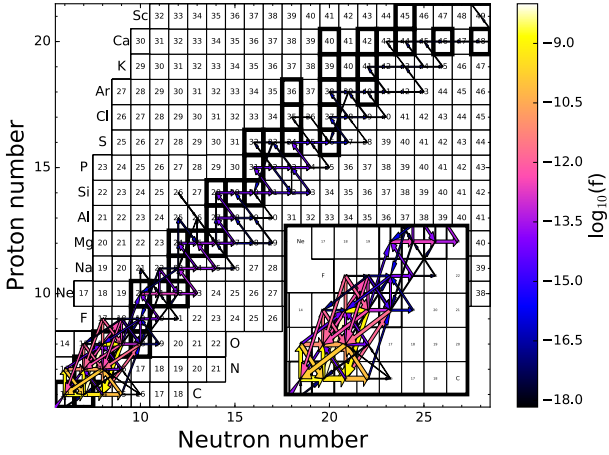

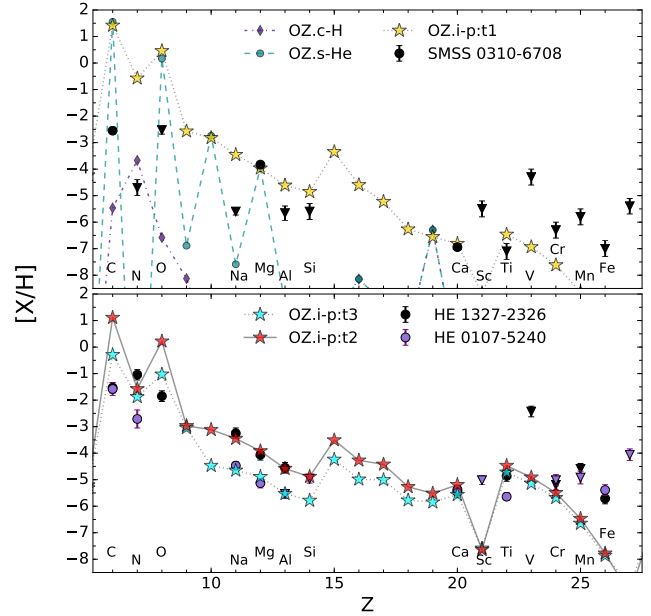

**Figure 3.** Nucleosynthesis fluxes showing the extent of the *i* process for the final model in run OZ.i-p:t1, $\log_{10}(\mathrm{f}) = \log_{10}(\mathrm{dY_i/dt})$.

**Figure 4. Top Panel:** Abundances of SMSS J0313-6708 with upper limits shown as triangles (Nordlander et al. 2017) compared to abundances from PPN runs OZ.c-H, OZ.s-He and OZ.i-p:t1 with dilution applied, see Section 3. **Bottom Panel:** Same for HE 1327-2326 and HE 1017-5240 after being diluted by factors $5 \times 10^{-3}$ and $1 \times 10^{-2}$ by mass, respectively.

The second scenario represents the case where He burning would have been able to advance further before the H/He mixing starts. The He-burning run OZ.s-He:+ adopts a higher temperature, but still within the range found in the He-burning shell (lower panel of Fig. 2) and runs for about 3.5 times longer than OZ.s-He. At this point the C, O and Mg are almost in the same proportions as in SMSS J0313-6708, similar to what is suggested by Maeder & Meynet (2015). The temperature and density were taken from the base of the He-burning shell just prior to the ingestion event. The OZ.s-He:+ case could be representative of a later H-ingestion event (into either the He burning core or shell) or a situation found in a model with different initial mass or different macroscopic mixing assumptions.

Each of the one-zone He-burning runs is followed by one or more one-zone models representing the H ingestion event. We add 1% H, by mass, to the output of the He-burning runs and renormalise all other isotopes. In these third one-zone models H burns rapidly in the $^{12}\mathrm{C}(p,\gamma)^{13}\mathrm{C}$ reaction, followed by $\beta$ decay and neutron release in the $^{13}\mathrm{C}(\alpha,n)^{16}\mathrm{O}$ reaction, exactly the same as in the one-zone *i*-process calculations by Dardelet et al. (2014). The resulting nucleosynthesis is also similar in that high neutron densities typical for *i* process are reached and the nucleosynthesis path in the chart of isotopes includes n-rich unstable isotopes (Fig. 3). The one-zone models representing the H-ingestion episode are therefore labelled OZ.i-p (Table 1) with OZ.i-p:+ being the H-ingestion run following OZ.s-He:+.

We finally assume that the products of nucleosynthesis would be diluted by either or both the stellar envelope, which the material would have travelled through to reach the surface of the star, and subsequently, the ISM. The relative amount of dilution from the envelope and ISM individually is not yet clear. To directly compare the abundances of SMSS J0313-6708 we dilute the material such that the amount of *i*-process material is 0.15% for run OZ.i-p:t1 and $10^{-5}$ for run OZ.i-p:+ and the remainder has the Big Bang abundance distribution. These numbers are chosen to fit the Mg abundance for OZ.i-p:t1 and C for OZ.i-p:+.

Fig. 4 shows the results of the core-H (OZ.c-H), shell-He (OZ.s-He) and finally the *i*-process run at time t1 (OZ.i-p:t1, top panel) after dilution. The neutron densities rise to $\approx 6 \times 10^{13}\ \mathrm{cm}^{-3}$ in both OZ.i-p:t1 and OZ.i-p:+ *i*-process runs. Run OZ.i-p:t1 has a C/Mg ratio much larger than observed because it reflects the beginning of He burning. The exact time for H ingestion is poorly constrained, and using input abundances from more complete He burning can yield the observed C/Mg ratio, as is the case in run OZ.i-p:+. In the latter case much more Mg is produced in He burning which is reflected by much greater required dilution to compare with observations. In order to reproduce the observed Ca abundance from the n-capture reactions a higher neutron exposure of $\tau = 3\ \mathrm{mbarn}^{-1}$ was realized in this case.

According to Keller et al. (2014), Bessell et al. (2015), and Takahashi et al. (2014) observed abundances of Ca in SMSS J0313-6708 can be produced during H burning via breakout reactions. The production of Ca in our models by this reaction channel is at least 1 dex lower than the observed Ca abundance in SMSS J0313-6708. Takahashi et al. (2014) report Ca production in H-shell burning at temperatures reaching $\log(T) = 8.66$ in models with masses initially in the range $80 - 140\,\mathrm{M}_\odot$. We do not find such high temperatures in any of our stellar evolution models, including tests with similar high and even higher initial mass. Pop III models of Limongi & Chieffi (2012) used by Marassi et al. (2014) are in better agreement with ours as they have similar H burning temperatures and do not produce appreciable Ca in quiescent burning phases.

In our simulations, Ca is primarily produced through n captures in *i*-process conditions in the form of $^{48}$Ca. The production site of this isotope has been a long-standing question in the nucleosynthesis community (Meyer et al. 1996). Previous scenarios to make $^{48}$Ca include anomalous CCSN conditions in parts of the ejecta (Hartmann et al. 1985), and the weak r-process (Wanajo et al. 2013).

It has been pointed out that H-ingestion events lead to Na production (Limongi & Chieffi 2012). Na is overproduced in the *i*-process simulation compared to the observed abundance in SMSS J0313-6708 by > 2.5dex. A preliminary exploration of several nuclear physics uncertainties have not offered an obvious pathway to change this result and it seems unlikely that 3D effects would fundamentally do so either. Interestingly, the [Na/Mg] ratio of SMSS J0313-6708 is indicative of a strong odd-even effect often seen in yields of core collapse supernova (Prantzos 2000), yet the upper limit of [Mg/Si] together with the low Al upper limit, and the high [Mg/Ca] at these low-metallicities can be accommodated by the *i*-process model. In our single-zone calculations Mg is produced in He burning. Na, Al, Si and Ca are produced during the H-ingestion *i* process phase. The α elements among these have two completely different nucleosynthetic origins.

We also compare our models with the next two most iron-poor stars known, HE 1327-2326 and HE 1017-5240. Fig. 4, bottom panel, shows the model OZ.i-p:t2 and OZ.i-p:t3 which are from the same run as OZ.i-p:t1, except at later times when the neutron exposure has further increased. For the three successive runs $\tau \approx 2.5, 4,$ and $8.75$ mbarn$^{-1}$. These simulations reproduce the observed sloping pattern of abundances and upper limits for Na, Mg, Al and Si very well. In this scenario, the composition of these CEMP-no stars could have been made by the contribution of a H-ingestion event, and at least one CCSN event contributing to the observed Fe-group elements. Interestingly, these models reproduce a similar decreasing abundance pattern from Ti, Cr to Mn that can be seen in some of the most iron-poor stars such as, CD-38°245, HE 1310-0536 or CS 22949-037 (Placco et al. 2016), which in our model is due to efficient absorption of neutrons around the magic number N = 28. This feature is often accompanied by the positive [Na/Mg] and [Mg/Al] ratios mentioned. Our preliminary tests show that both of these sloping abundance patterns for Na-Al and Ti-Mn can also be found if the initial metal abundance is hyper metal-poor instead of Pop III. The [Na/Mg] ratio could be as small as ≈ −0.3 dex in these conditions. But the presence Fe-group n-capture seeds, or even a too high neutron exposure with Pop III initial abundance, would lead to an efficient production of trans-iron elements with the *i*-process abundance signature, leading not to CEMP-no but to CEMP-i (or CEMP-r/s) star abundance signatures (Dardelet et al. 2014; Herwig et al. 2011; Hampel et al. 2016).

## 4 CONCLUSION

We compared Pop III *i*-process models with the three most iron-poor stars SMSS J0313-6708, HE 1327-2326 and HE 1017-5240. The Pop III *i*-process models are based on massive Pop III stellar evolution models which undergo a highly energetic H-ingestion event during He shell burning. MLT description of convection breaks down in this situation, and 1D models are therefore unreliable. 3D simulations are required. In the meantime we therefore use a series of single-zone nucleosynthesis calculations in an initial attempt to identify the nucleosynthetic signature of Pop III *i* process in isolation. The general abundance distribution of SMSS J0313-6708 is reproduced by our models, but Na is overproduced by > 2.5 dex, which likely rules this scenario out for SMSS J0313-6708. When comparing our models to HE 1327-2326 and HE 1017-5240, the high [Na/Mg] and [Mg/Al] ratios, which are difficult to reconcile with metal poor CCSN models, are well reproduced, and Pop III *i* process becomes a promising candidate for at least a portion of the nucleosynthetic material of which these stars are made. Our models are able to recover the low observed [Ca/Mg].

Many details in this scenario remain uncertain and will be subject to further investigation. The greatest areas of uncertainty are the convective nature of the event in 3D, the single zone-treatment of the nucleosynthesis and the uncertain nuclear physics data of n-rich, unstable light elements.

## ACKNOWLEDGEMENTS

This work was supported by NSF grant PHY-1430152 (JINA Center for the Evolution of the Elements), Compute Canada and NSERC. We acknowledge contributions from Sam Jones to the NuGrid codes and to earlier explorations of Pop III massive star models.